# ANALYSIS OF $M_2/M_2/1/R, N$ QUEUING MODEL FOR MULTIMEDIA OVER 3.5G WIRELESS NETWORK DOWNLINK

SULEIMAN Y. YERIMA AND KHALID AL-BEGAIN

Mobile Computing, Communications and Networking RG
Faculty of Advanced Technology, University of Glamorgan
Pontypridd (Cardiff) CF37 1DL, Wales, UK
E-mail: {syerima,kbegain}@glam.ac.uk

**ABSTRACT**

Analysis of an $M_2/M_2/1/R$, N queuing model for the multimedia transmission over HSDPA/3.5G downlink is presented. The queue models the downlink buffer with source multimedia traffic streams comprising two classes of flows: real-time and non real-time. Time priority is accorded to the real-time flows while the non real-time flows are given buffer space priority. An analytic evaluation of the impact of varying the buffer partition threshold on the QoS performance of both classes of customers is undertaken. The results are validated with a discrete event simulation model developed in C language. Finally, a cost function for the joint optimization of the traffic QoS parameters is derived.

## INTRODUCTION

3G WCDMA mobile cellular networks have been widely deployed across the globe to support higher data rate applications and services that previous cellular networks were unable to provide. In recent times, new standards and recommendations for the enhancement to 3G to meet insatiable demand for new wireless broadband and broadcast services have been introduced. In early 2002, Release 5 introduced improved support for downlink packet data over WCDMA networks in the form of High Speed Downlink Packet Access (HSDPA) technology otherwise referred to as 3.5G (Van den berg et al. 2005).

In addition to significantly reducing downlink latency, HSDPA enables peak data rates of 14.4 Mbps and a three-fold capacity increase in WCDMA networks. The enhancement to WCDMA is achieved mainly through the following techniques: Link adaptation, Hybrid ARQ, and Fast Scheduling. To meet the demand on low latency and rapid resource (re)allocation, the enhancing functionalities have been located in the Node B, the base station, as part of additions to the WCDMA MAC layer. Furthermore, a shorter transmission time interval (TTI) of 2ms, is employed which reduces overall delay and improves tracking of fast channel variations exploited by the link adaptation and channel-dependent fast scheduling. See (3GPP 2001; Kolding et al. 2003; Parkvall et al. 2006) for further details on HSDPA.

With the improved support for broadband and broadcast services in WCDMA networks, the growth in demand for multimedia-based applications and services is set to escalate. In a HSDPA network carrying downlink multimedia traffic destined for a user terminal in one of its cells, QoS performance gains could be achieved from differentiated consideration of the several flows comprising the multimedia streams as they traverse the Node-B downlink buffer. Based upon this idea, a novel queuing model for performance evaluation of multimedia traffic over 3.5G wireless downlink was introduced in (Al-Begain et al. 2005). In both (Al-Begain et al. 2005) and (Al-Begain et al. 2006), the model has been investigated analytically with BMAP/PH/1/R,N and $M2/M2/1/N,\infty$ queues respectively. While in (Yerima and Al-Begain, 2006), a comparative analysis of the downlink buffer partitioning schemes based on the model was carried out.

This paper extends the work presented in (Yerima and Al-Begain, 2006) by further investigation of the impact of varying the buffer partitioning threshold on the QoS parameters of the multimedia traffic, by means of analytical modelling. The results obtained are then validated using Discrete Event Simulation of the system in C language. Additionally, a cost function for joint optimization of the investigated QoS performance metrics is derived, and, finally it is shown that for a given set of traffic and system parameters an optimum buffer partition threshold is obtainable from the cost function.

The rest of the paper is organized as follows. Section 2 describes the conceptual model of the system. Section 3 is devoted to the analytical evaluation of the model, while numerical results of the experiments are presented in section 4. Finally, the concluding remarks are given in section 5.

## MODEL DESCRIPTION

A single-server priority queue with a finite capacity T = N + R, is considered for the buffer model. From the multimedia source, two classes of traffic, real-time (RT) and non real-time (NRT), arrive according to the stationary Poisson process with mean rates $\lambda_{rt}$ and $\lambda_{nrt}$ respectively. Service times are assumed to be exponentially distributed with mean rates $\mu_{rt}$ and $\mu_{nrt}$ with priority service given to real-time customers whenever both types are present in the buffer. A non pre-emptive priority access to service is accorded to the real-time flows. Scheduling of service is FIFO (First-In-First-Out) for both customers. The maximum numbers of real-time customers that can be admitted into the buffer is restricted to R, R > 0, and on arrival, are placed ahead of the non real-time customers to be scheduled for transmission.

Thus, real-time customers have 'time priority' over non real-time customers, but, due to restriction of the real-time customers' access to the buffer, non real-time customers have a 'space priority'. Limiting the number of admissible real-time customers to a constant, R, effectively partitions the buffer, with R representing the threshold as shown in Figure 1. Thus, the QoS performance metrics such packet loss probability and mean waiting time, of both customers are determined by R, the threshold position which is also the number of admissible real-time customers. Dynamic control of the QoS performance metrics to meet optimization goals or application requirements can potentially be achieved by adaptively varying the threshold R.

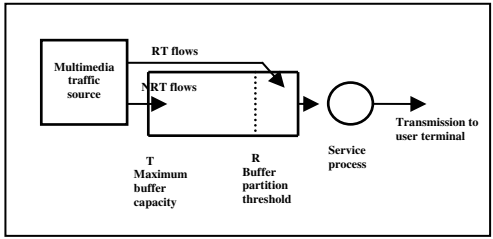

Figure 1: HSDPA Downlink Buffer Model

## ANALYTICAL EVALUATION

The system state is described by the stochastic process $S(t) = (R(t); N(t))$, $t \geq 0$; where $R(t)$ is the number of real-time customers and $N(t)$ represents the number of non real-time customers. All random variables are exponentially distributed and hence the underlying stochastic process is a two-dimensional continuous-time Markov chain (CTMC) with finite state space as depicted in Figure 2. The steady-state probabilities, $P(i,j)$, of the system states are defined by:

$$P(i,j) = \lim_{t \to \infty} P(R(t) = i, N(t) = j), \quad i = \overline{0,R}, \; j = \overline{0,N}$$

If the steady-state probability vector of all the possible states $P(i,j)$ of the CTMC is denoted by **P,** then the steady-state probabilities can be obtained by solving the following set of equations:

$$\mathbf{PG=0} \quad \text{and} \quad \mathbf{Pe=1} \qquad (1)$$

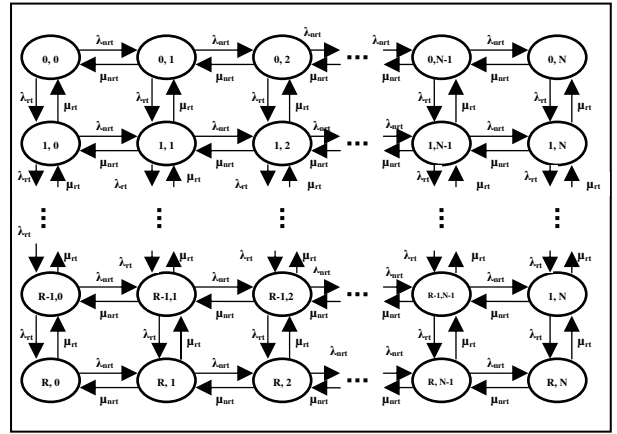

Figure 2: CTMC for the $M_2/M_2/1/R$, N queuing system

Where **G** is the transition probability matrix and **e** is a column vector of the appropriate dimension consisting of ones. If we denote by $P_{ij \to i'j'}$ the steady-state probability of transition from a given state $S = (i, j)$ to another state $S' = (i', j')$, then from observation of the CTMC, the entries of the matrix **G** are defined by:

$P_{ij \to i'j'} =$

$$\begin{cases} \lambda_{nrt} & i' = i, \; j' = j+1 \quad j \neq N \\ \lambda_{rt} & i' = i+1, \; j' = j, \quad i \neq N \\ \mu_{nrt} & i' = i, \; j' = j-1, \quad j \neq 0 \\ \mu_{rt} & i' = i-1, \; j' = j, \quad i \neq 0 \\ (\lambda_{rt} + \lambda_{nrt}) - (\mu_{nrt} + \mu_{rt}) & i' = i = 0, \; j' = j = 0 \\ \lambda_{rt} - \mu_{rt} & i' = i = 0, \; j' = j \; (1 \leq j \leq N-1) \\ (\lambda_{rt} + \mu_{nrt}) - (\lambda_{nrt} + \mu_{rt}) & i = i = 0, \; j' = j = N \\ \lambda_{nrt} - \mu_{nrt} & i' = i \; (1 \leq i \leq R-1), \; j' = j = 0 \\ (\lambda_{nrt} + \mu_{rt}) - (\lambda_{rt} + \mu_{nrt}) & i' = i = R, \; j' = j = 0 \\ \mu_{rt} - \lambda_{rt} & i' = i = R, \; j' = j \; (1 \leq j \leq N-1) \\ (\mu_{nrt} + \mu_{rt}) - (\lambda_{rt} + \lambda_{nrt}) & i' = i = R, \; j = j = N \\ \mu_{nrt} - \lambda_{nrt} & i' = i \; (1 \leq i \leq R-1), \; j' = j = N \\ 0, & \text{otherwise} \end{cases}$$

(2)

The system performance measures are calculated from the following set of equations. Mean number of real-time customers is given by:

$$N_{rt} = \sum_{i=0}^{R} \sum_{j=0}^{N} i \, P(i,j) \qquad (3)$$

Mean number of non real-time customers is given by:

$$N_{nrt} = \sum_{i=0}^{R} \sum_{j=0}^{N} j \, P(i,j) \qquad (4)$$

Loss probability of real time customers is given by:

$$L_{rt} = \sum_{j=0}^{N} P(R,j) \qquad (5)$$

Loss probability of non real time customers is given by:

$$L_{nrt} = \sum_{i=0}^{R} P(i,N) \tag{6}$$

With the equations (3) – (6), the mean delay or waiting time can be calculated using Little's theorem (Bolch et al. 1998) thus:

Mean delay for real-time customers is given by:

$$D_{rt} = \frac{N_{rt}}{\lambda_{rt} \times (1 - L_{rt})} \tag{7}$$

Similarly, mean delay for non real-time customers is given by:

$$D_{nrt} = \frac{N_{nrt}}{\lambda_{nrt} \times (1 - L_{nrt})} \tag{8}$$

**Joint Optimization of the System QoS Parameters**

As mentioned earlier, the threshold of the buffer can be varied to meet QoS requirements or optimize the QoS parameters according to some given criteria. In (Al-Begain et al. 2006), an economic criterion γ called the Weighted Grade of Service (WGoS) was formulated for the investigated queuing network model. An adapted form of the WGoS criterion suited to the $M_2/M_2/1/R, N$ buffer queuing model being investigated in this paper is assumed to be of the following form:

$$\gamma = \frac{\lambda_{rt}}{\lambda_{rt} + \lambda_{nrt}} [CL_{rt} \times L_{rt} + (1 - L_{rt}) \times CD_{rt} \times D_{rt}] \\ + \frac{\lambda_{nrt}}{\lambda_{rt} + \lambda_{nrt}} [CL_{nrt} \times L_{nrt} + (1 - L_{nrt}) \times CD_{nrt} \times D_{nrt}] \tag{9}$$

Where $CD_{rt}$ is the penalty for the mean delay of RT customers; $CD_{nrt}$ is the penalty for the mean delay of NRT customers. Likewise, $CL_{rt}$ is the penalty for the loss of RT customers while $CL_{nrt}$ is the cost penalty for the loss of NRT customers. We use γ, the WGoS function, to determine the optimum operating threshold position for a given set of traffic and system parameters, in the next section.

**NUMERICAL RESULTS**

In this section, the results of two sets of experiments are presented. The first investigates the impact of the buffer partition threshold on traffic performance. The second employs the optimization cost function, γ, to determine the optimum buffer partition threshold for a given set of traffic parameters.

**Impact of Buffer Partition Threshold on Traffic Performance**

From the analytical model developed earlier, we study the effect of varying the buffer partition threshold R, on traffic performance. Performance measures are obtained from the formulae shown earlier, using the analytical modelling tool MOSEL (Al-Begin et al. 2002; Beutel 2003). In order to validate the analytical study, an equivalent simulation model was developed in C language and the results from both models are illustrated side-by-side. Figures 3 to 6 show the results of the first scenario assuming the following parameters: $\lambda_{nrt} = 6$, $\mu_{nrt} = 10$, $\mu_{rt} = 20$. The buffer partition threshold R is varied from 2 to 16 whilst buffer capacity T= R+N is fixed at 20. The loss and delay performances for both classes of traffic are shown for $\lambda_{rt}$ = 2, 12, and 18 corresponding to low, medium and high RT traffic loads respectively.

In Figure 3 we see that the loss probability of NRT traffic increases with R which is because less buffer space becomes available to NRT packets as R increases. Notice also the deteriorating loss performance with increase in $\lambda_{rt}$ for a given buffer partition threshold, which is as a result of the priority access to service enjoyed by the RT traffic. Figure 4 shows how R affects the mean NRT delay. It is clear that at medium and high $\lambda_{rt}$ load, the mean delay increases, peaks and then decreases again as R is increased. The mean NRT delay increases at first because more RT traffic is buffered as R increases. The drop in the mean delay with further increase in R can be attributed to less NRT traffic being retained in the buffer, as the corresponding NRT buffer space N decreases correspondingly. Figure 5 shows RT loss probability dropping with increase in R as expected, while in Figure 6 mean RT delay is seen to increase with R and vice versa. Notice also from Figures 4 to 6 that low intensity of RT traffic seems to have only marginal effect on traffic performance for this scenario.

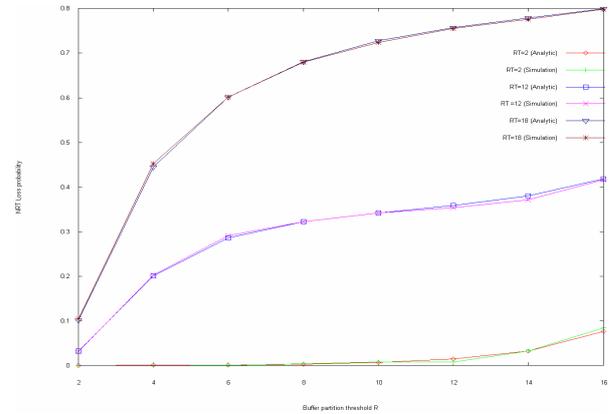

Figure 3: NRT loss Vs R for $\lambda_{rt}$ = 2, 12, and 18

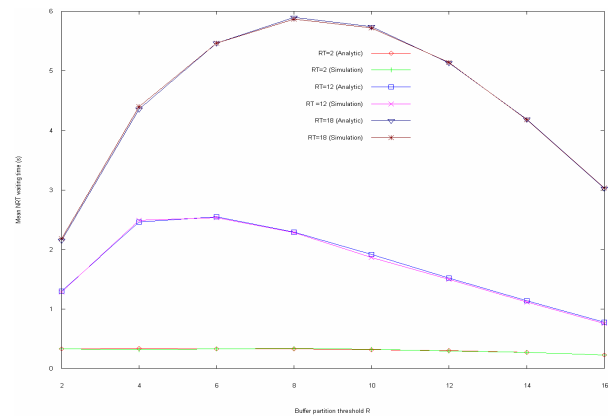

Figure 4: NRT delay Vs R for $\lambda_{rt}$ = 2, 12, and 18

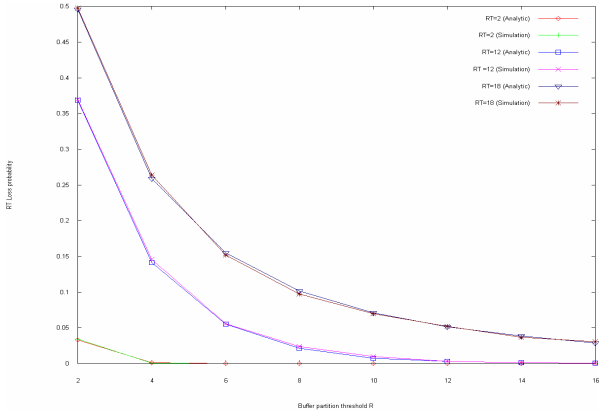

Figure5: RT loss Vs R for $\lambda_{rt}$ = 2, 12, and 18

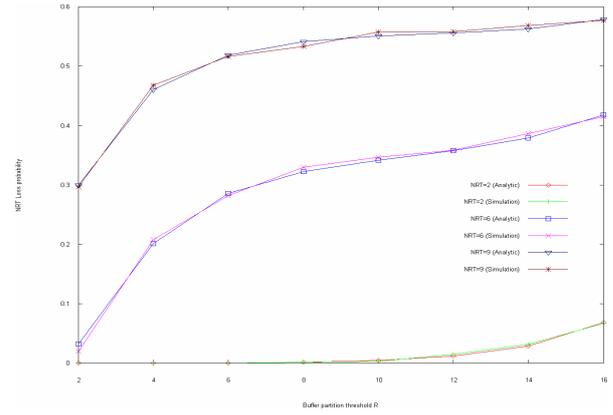

Figure7: NRT loss Vs R for $\lambda_{nrt}$ = 2, 6, and 9

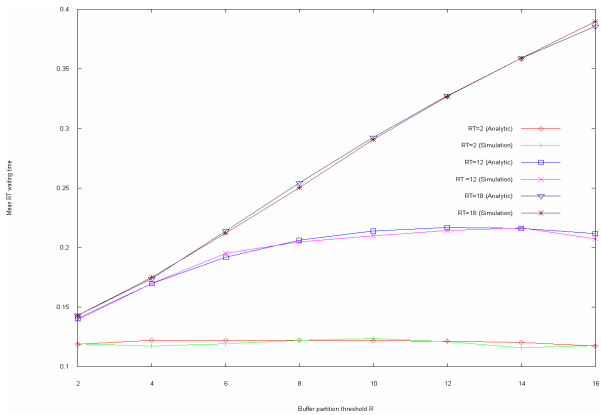

Figure 6: RT delay Vs R for $\lambda_{rt}$ = 2, 12, and 18

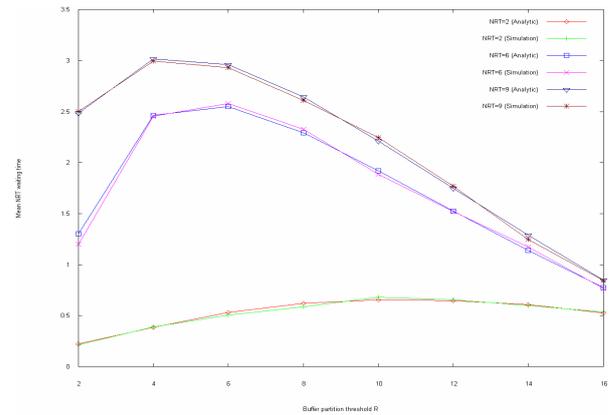

Figure8: NRT delay Vs R for $\lambda_{nrt}$ = 2, 6, and 9

Similarly, Figures 7 to 11 illustrate the results of the second scenario investigated with the following parameters: $\lambda_{rt}$ = 12, $\mu_{rt}$ = 20, $\mu_{nrt}$ = 10 and T=20. R is again varied from 2 to 16 and $\lambda_{nrt}$ assumes the values 2, 6, and 9 corresponding to low, medium and high NRT traffic. Figure 7 shows similarity to figure 3, i.e. NRT loss probability increasing with larger R and also with higher NRT traffic. In Figure 8, the mean NRT delay increases, peaks and then drops again with increase in R as in Figure 4. Thus, the same reasons given earlier that account for the behavior in Figure 4 also apply here.

Finally, the RT traffic performance for the second scenario is depicted in Figures 9 and 10. RT loss probability is seen to decrease with R as expected (c.f. Figure 9), but, increasing NRT intensity from medium to high loads seem to have very little effect on RT loss. This could also be attributed to the prioritized access to service accorded to RT traffic which to some extent shields it from effects of λnrt variation. From Figure 10 we also see that as R increases so does mean RT delay. Again, due to RT service priority access, high and medium NRT have only marginal effect on the mean RT delay performance. The results of the experiments suggest that varying the buffer partition threshold affect the QoS performance metrics of the multimedia traffic differently and, thus, an optimum threshold can be found by trading off the QoS performance metrics against each other.

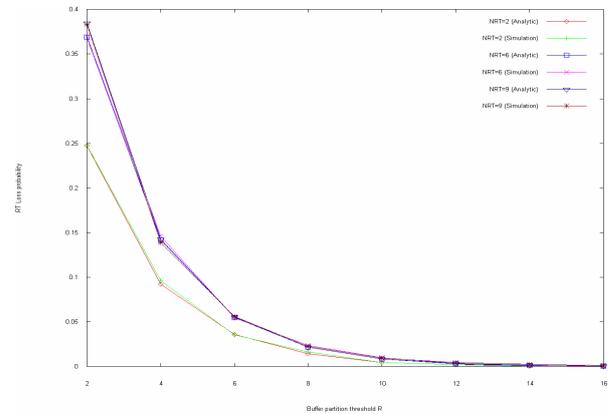

Figure9: RT loss Vs R for $\lambda_{nrt}$ = 2, 6, and 9

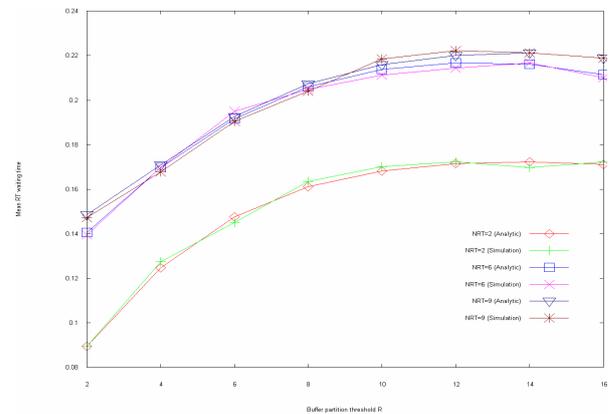

Figure10: RT delay Vs R for $\lambda_{nrt}$ = 2, 6, and 9

## Deriving Optimum Buffer Partition Threshold from Traffic Parameters

From the WGoS equation (9), it is clear that the optimum buffer capacity threshold can be determined for a given set of traffic parameters since the performance metrics are dependent on those parameters. For this experiment, the cost values are taken as follows: $CL_{rt}$ = 300, $CL_{nrt}$ =50; $CD_{rt}$=1000, and $CD_{nrt}$=1. From Figure 11 we see that for $\lambda_{rt}$ = 12, $\lambda_{rt}$ = 18 the optimum buffer capacity threshold, satisfying the WGoS economic criterion with the above given cost parameters, is the minimum value of 3. Other traffic parameters are taken as: $\lambda_{nrt}$ = 6, $\mu_{nrt}$ = 10, $\mu_{rt}$ = 20 and T = 20 respectively. Likewise, Figure 12 shows that for both $\lambda_{nrt}$ = 6, $\lambda_{nrt}$ = 9 and traffic parameters: $\lambda_{rt}$ = 12, $\mu_{rt}$ = 20, $\mu_{nrt}$ = 10, T=20, the buffer threshold which minimizes γ, the WGoS, is 3.

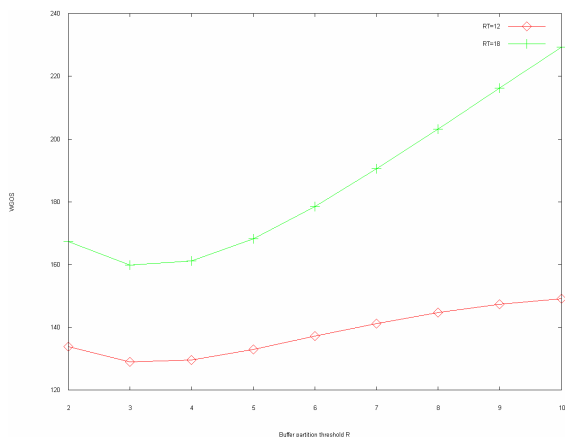

Figure 11: WGoS Vs R for $\lambda_{rt}$ = 12, and 18

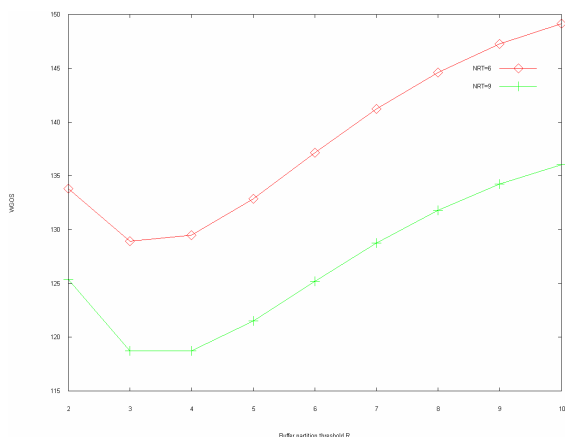

Figure 12: WGoS Vs R for $\lambda_{nrt}$ = 2, 6, and 9

## CONCLUDING REMARKS

In this paper analysis of the $M_2/M_2/1/R$, N queuing model for multimedia downlink transmission in 3.5G is presented. The results of the experiments illustrate clearly the impact of varying the downlink buffer partition threshold on the traffic performance at various load levels. Results of both analytic evaluation and simulation are shown to be in very close agreement. In addition, an economic criterion i.e. the WGoS, for the joint optimization of the QoS parameters is derived for the model. Finally, it is shown that for a given set of traffic and system parameters an optimum buffer partition threshold which jointly optimizes the multimedia traffic QoS performance can be obtained from the WGoS. In our future work, other arrival and service distributions will be considered for the queuing model. Furthermore, we aim to incorporate the model into holistic simulated HSDPA communication scenarios and then carry out performance evaluation.